\documentclass[doublecol]{epl2}

\usepackage{amsmath}
\usepackage{amsfonts}
\usepackage{amssymb}
\usepackage{graphicx}
\usepackage{dcolumn}
\usepackage{bm}
\usepackage{textcomp}
\usepackage{footnote}

\title{Universal Scaling Property of System Approaching Equilibrium}
\author{P. Barat \inst{1} \and A. Giri \inst{1}\and M. Bhattacharya \inst{1}\and Nilangshu K. Das\inst{1} \and A. Dutta\inst{2}}
\shortauthor{P. Barat \etal}
\institute{                    
  \inst{1} Variable Energy Cyclotron Centre,1/AF Bidhannagar, Kolkata 700064, India\\
  \inst{2} S.N. Bose National Centre for Basic Sciences, Block JD, Salt Lake, Kolkata 700098, India
}

\abstract{
In this Letter we show that the diffusion kinetics of kinetic energy among the atoms in non-equilibrium crystalline systems 
follows universal scaling relation and obey Levy-walk properties. This scaling relation is found to be valid for systems no
matter how far they are driven out of equilibrium.}

\begin{document}
\maketitle

             Non-equilibrium systems are ubiquitous in nature. Equilibrium systems are ideal and can only be achieved in the 
laboratory. Attempts have been made to understand the dynamics of non-equilibrium systems by linear response theory \cite{1} albeit
 its domains of validity are restricted to the linear response regime. There exists no general formalism to deal 
with systems that are far from equilibrium. For a system out of equilibrium, the probability of a given microstate evolves 
continuously with time. In the long time limit the system reaches a stationary state in which the probability measure over the 
configuration space converges to a constant distribution. Non-equilibrium systems display fluctuations which are less sensitive 
to the conditions of the surroundings and carry information about the dynamics of its present state. In the last decade certain 
general relations have been discovered which are valid for non-equilibrium systems and are independent of how far the system is 
driven out of equilibrium. These results include the Jarzynski equality \cite{2,3} and the fluctuation theorems 
\cite{4,5,6,7,8,9,10,11,12}. They have been verified for a variety of systems theoretically as well as experimentally 
\cite{13,14,15,16,17}. After the work by Crooks \cite{7} and Seifert \cite{8}, it is now understood that many of these relations
 are closely related to the path probability of the system’s trajectory. In the absence of any general theory to understand the
 dynamics of a system far from equilibrium, one approach will be to take a simple but nontrivial model system and try to 
understand its dynamics when it approaches equilibrium from its non-equilibrium state. The general dynamical behavior of a 
non-equilibrium system will consist of superposition of various dynamics on well speared time scales which compel several 
thermodynamic parameters of the system like heat, work, internal energy, kinetic energy of the particles of the
 system to fluctuate. To understand the general features of the dynamics one has to apply statistical analysis to these
fluctuations. In this Letter an attempt has been made to study a universal scaling relation for systems driven far from equilibrium 
by analyzing its fluctuation properties. The non-equilibrium systems of study are generated by molecular dynamics simulation 
technique.

Dynamical systems arising from diversified disciplines of science can be quantified in a unified way from their scale invariance properties. Scafetta 
$et$ $al.$ \cite{18} introduced two complementary scaling analysis methods: Diffusion Entropy Analysis (DEA) and Finite Variance Scaling 
Method (FVSM) to evaluate correct scaling that prevails in complex dynamical systems. DEA studies the scaling exponent $\delta $ 
of the probability distribution instead of its moment, based on the evaluation 
of Shannon entropy $S(t)$ of a time series $\{\xi_i\}$. The DEA has been successfully applied to time series of 
different kinds of dynamical systems \cite{19}. In DEA analysis, the numbers in the time series, $\{\xi_i\}$, are 
thought to be the fluctuations of a diffusion trajectory \cite{19} arising from the probability density function (PDF), $p(x,t)$, of 
the corresponding diffusion process. Here $x$ denotes the variable collecting the fluctuations and is referred as the diffusion 
variable. The scaling property of $p(x,t)$ is evaluated by means of the sub trajectories $ x_n(t)=\sum_{i=0}^{t}\xi_{i+n}$ 
 with n = 0,1,....etc. If the scaling condition $ p(x,t)=t^{-\delta}F(x t^{-\delta})$
  holds true, it is easy to prove that the entropy $S(t)$ is given by $ S(t)=-\int^{+\infty}_{-\infty}p(x,t)ln[p(x,t)]dx=A+\delta ln(t)$,
 where $A$ is a constant. This indicates that in the case of a
 diffusion process with a scaling PDF, its entropy $S(t)$ increases linearly with $ln(t)$. One can also examine the scaling properties
 of the second moment of the same process by FVSM. One version of FVSM is the Standard Deviation Analysis (SDA) \cite{19}, which is 
based on the evaluation of the standard deviation $D(t)$ of the variable $x$, and yields $ D(t)= \left[\langle x^2;t\rangle-\langle x;t\rangle^2 \right]^\frac{1}{2}
\propto t^H$ \cite{19}. The exponent $H$ is interpreted as the scaling exponent. For random noise 
with finite variance, the diffusion distribution $p(x,t)$ will converge, according to the central limit theorem, to a Gaussian
 distribution with $H =\delta = 0.5$. If  $H \neq \delta $, the scaling represents anomalous behavior. Levy-walk is a kind of anomalus diffusion
 which is obtained by generalizing the central limit theorem \cite{20}. In this
 particular kind of diffusion process the scaling exponents $H$ and $\delta $ are found to obey the 
relation $\delta=(3-2H)^{-1} $ instead of being 
equal \cite{19}.
\begin{savenotes}
\begin{table*}[!ht]
\caption{ The parameters used for the MD simulations.}
\label{table1}
\begin{center}
\begin{tabular}{lcccccccc}
\hline
\hline
\\
\bf{Material} & \bf{Si} & \bf{Ge} & \bf{Cu} & \bf{Solid Ar} & \bf{Mo} & \bf{Fe}&\bf{Al}\\\\
\hline
\\
Simulation cell size (in unit cell)  & $25^3$ & $15^3$  & $20^3$ &$20^3$ &$25^3$ & $25^3$ & $20^3$&\\\\
Number of atoms in simulation cell & 125000&27000 &32000 &32000 & 31250& 31250&32000&\\\\
Simulation time step (fs) &0.25 & 0.5&0.5 &2.0& 0.5&0.5 &0.5&\\\\
Nature of inter-atomic potential used &SW\footnote{SW (Stillinger-Weber)} &SW &EAM\footnote{ EAM (Embedded Atom Model)} &LJ\footnote{LJ (Lennard-Jones)} &FS\footnote{FS (Finnis-Sinclair)} &FS &GLUE\\\\
Total number of data taken &6000 &3000 & 3000&10000 & 3000& 3000&3000&\\\\
Radius `r' (\AA) & 8.8250&9.0809 & 5.4766&7.9952 &6.2944 &5.7330 &6.1566&\\\\
No of atoms in the sphere of radius `r'& 150&147 &55 &55 &63 &63 &55&\\\\
\hline
\hline
\end{tabular}
\end{center}
\end{table*}
\end{savenotes}

\begin{figure}[!ht]
\includegraphics[width= 0.445\textwidth]{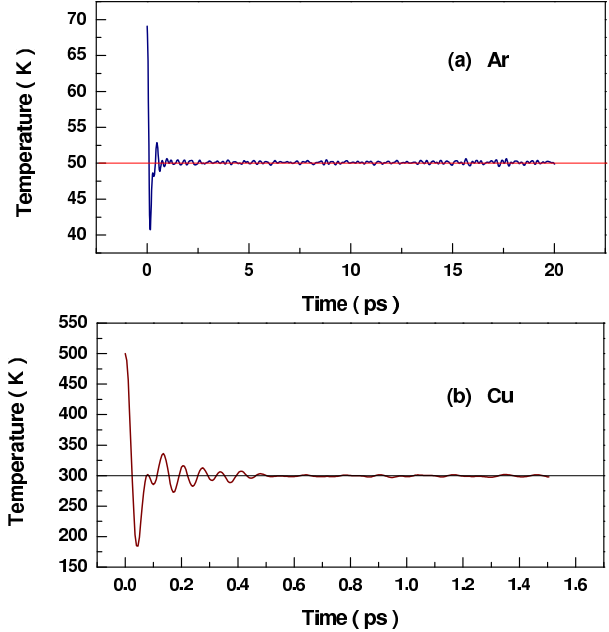}
\vspace{-10pt}
\caption{Variations of temperature (equivalent to KE ) with time of the systems of (a) Ar and (b) Cu leading to the equilibration}
\vspace{-5pt}
\label{figg1}
\end{figure}

To generate the non-equilibrium model system we have utilized the molecular dynamics (MD) simulation technique in an innovative 
way. Depending on the interaction potential model, a typical MD simulation computes the trajectories of atoms in a system by
solving Newton’s equations of motion numerically. In our study, 
the simulations have been performed for elements having different crystal structures like Silicon (Si) and Germanium (Ge) with
 diamond cubic structure, Iron (Fe) and Molybdenum (Mo) with body centered cubic structure and Copper (Cu), Aluminium (Al) and solid Argon (Ar)
 having face centered cubic structure. Thus, a variety of interaction potential models have been employed (refer Table~\ref{table1}) in the
 simulations. For all these elements, periodic boundary conditions are imposed along three directions (x,y,z) in the simulation 
cell. Initially, all the systems are equilibrated using constant energy and volume ensemble (NVE) at 100 K for 2.5 ps except for
 Ar, where the equilibration run is performed at 30K for 10 ps. The kinetic energy in the equilibrium state will follow Boltzmann
 distribution.  The period of time for equilibration is chosen to be sufficient to bring the total energy of the system to divide
 approximately equally between kinetic energy (KE) and potential energy (PE). Details of the parameters used in the simulations like simulation cell size, number of atoms
 taken in the simulation cell, the time steps after which simulation data are recorded and the nature of the inter-atomic potentials used for the elements under study are given in Table I. To generate the non-equilibrium state, the three
 components of velocities of individual atom in the simulation cell were changed to random values such that  the instantaneous
 KE of the atom increases however the average temperature of the system corresponds to T = 500 K (T = 70 K for Ar). The atoms
 in the simulation cell are then allowed to equilibrate. \revision{After sufficiently long period of time, the system is observed to attain equilibrium temperature of 300 K (50 K for Ar) and the average kinetic energy becomes equal to the average potential energy of the respective systems. The variations of temperature (equivalent to KE) with time of the systems of Ar and Cu leading to equilibration are shown in Fig.~\ref{figg1}.  Atomistic simulation data are recorded for each time step for the entire equilibration time to study the dynamic nature of the equilibration process.}
The molecular dynamics simulations are carried out using MD++ simulation package \cite{21}.

\begin{figure}[!ht]
\includegraphics[width= 0.45\textwidth]{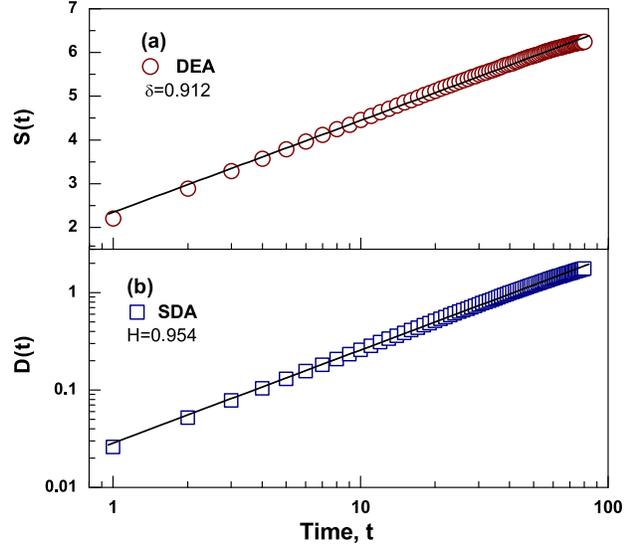}
\vspace{-10pt}
\caption{(a) DEA and (b) SDA of the variation of kinetic energy against time data obtained from a Copper atom in an ensemble when 
brought to a non-equilibrium state at 500 K from 100 K and allowed to equilibrate at 300 K}
\vspace{-5pt}
\label{figg2}
\end{figure}

Few atoms in the simulation cell are identified in a sphere of radius `r' whose centre is chosen to be almost at the middle of 
the cell, far away from the surface. The values of `r' and the number of atoms in the sphere for each simulation are given
 in Table~\ref{table1}. The KE in the consecutive time steps for the atoms in the sphere are used for further analysis. In these simulations
 the atoms in the simulation cell are kept isolated from the environment and hence there was no dissipation of energy, only there
 is a redistribution of KE and PE among the atoms. DEA and SDA analyses \cite{22} were performed for the time series data of the KE 
for an individual atom. Typical plots of DEA and SDA analysis from which  $\delta $ and $H$ are calculated for a Copper atom are shown in
 Fig.~\ref{figg2}.
 The slopes of the curves evaluate  $\delta $ and $H$. DEA and SDA analyses were performed for ten atoms randomly chosen from the
 atoms confined in the sphere of radius `r' and the average of values of  $\delta $ and $H$ are given in Table~\ref{table2}. Simulation 
cell sizes, number of atoms in the cell, sampling time, the crystal structure, and the nature of the interaction potential were
 varied to find the exact nature of the scaling.

\begin{table*}[!ht]
\caption{Mean values of the scaling exponents $\delta$ and H obtained from ten atoms chosen randomly in the shell of radius `r'.  Exponents $\delta$ and $H$ are evaluated from the time series of the evolution of KE for a single atom.}
\label{table2} 
\begin{center}
\begin{tabular}{lccccc}
\hline
\hline
\\
\bf{Material} & \bf{Non-equilibrium  } & \bf{$\delta$} & \bf{$H$} & \bf{$\left[\left(\delta-\frac{1}{3-2H}
\right)/\delta\right]\times100$} \\
&\bf{Temperature(K)}&&&\\\\
\hline
\\
Si&500&0.953\textpm0.016&0.968\textpm0.008&1.380&\\\\
Si&800&0.898\textpm0.025&0.901\textpm0.025&7.046&\\\\
Si ( single atom )&500&0.943&0.970&0.042&\\\\
Ge&500&0.930\textpm0.021&0.953\textpm0.026&1.712&\\\\
Ge&800&0.930\textpm0.020&0.960\textpm0.015&0.438&\\\\
Ar&70&0.968\textpm0.010&0.968\textpm0.005&2.908&\\\\
Ar ( single atom )&70&0.963&0.972&1.665&\\\\
Cu&500&0.920\textpm0.020&0.955\textpm0.022&0.279&\\\\
Cu ( single atom )&500&0.935&0.978&2.444&\\\\
Fe&500&0.925\textpm0.018&0.964\textpm0.022&0.847&\\\\
Fe ( single atom )&500&0.959&0.974&0.879&\\\\
Mo&500&0.935\textpm0.016&0.964\textpm0.021&0.231&\\\\
Al&500&0.935\textpm 0.008&0.966\textpm 0.013&0.142&\\\\
\hline
\hline
\end{tabular}
\end{center}
\end{table*}

To understand the process of equilibration when a single excited atom interacts with an ensemble of atoms that are in equilibrium,
 another kind of exercise was performed by MD simulation.  In this case from the equilibrated ensembles of Si, Cu, Fe at 300 K and 
solid Ar at 50 K one atom in the middle of the cell was excited to a KE corresponding to 500 K and 70 K respectively.  The excited 
atom is then allowed to equilibrate and the KE possesses by the atom at different time steps was monitored. In these simulations
 the energy of the simulation cells was also a constant of motion. All the equilibrated atoms in the cell form a heat bath and
 the excited atom equilibrate by interacting with this heat bath by sharing its excess KE.  The time series thus obtained of
 the KE of the excited atom between consecutive time steps was used to understand the diffusion process of KE and the nature 
of the dynamical process responsible to bring the atom to the equilibrium state. The values of $\delta $ and $H$ obtained by DEA and SDA
 analysis from the time series are given in Table~\ref{table2}.

\begin{figure}[!ht]
\includegraphics[width= 0.47\textwidth,clip=true ]{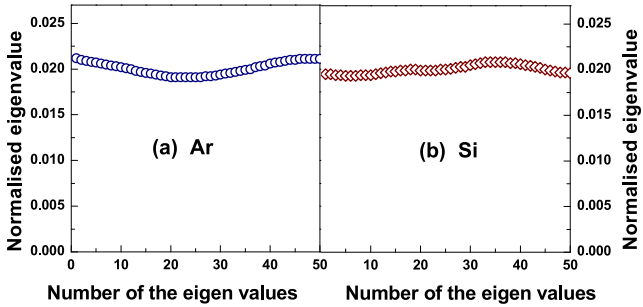}
\caption{Eigen value spectra of the covariance matrices constructed from the time series of the fluctuating forced fields 
experienced by (a) an Argon and (b) a Silicon atom when brought to a non-equilibrium state at 500 K from 100 K and allowed to equilibrate 
at 300 K}
\vspace{-8pt}
\label{figg4}
\end{figure}

Atoms in the simulation cell, when brought to a non-equilibrium state by enhancing their KE, try to redistribute their excess 
KE by the process of diffusion. This process of redistribution of KE among the atoms changes their mean positions of vibration
 and consequently their PE. Here all the atoms in the cell participate together in the diffusion process. Thus by analyzing the 
variation of the KE with time for a single atom in the cell one can account for the modality of this diffusion process and the
 nature of the equilibration. The change in the KE for an atom in the $i^{th}$ time step is given by 
$\Delta E=\left(\frac{2E_i}{m}\right)^\frac{1}{2}F_i\Delta t $ where $F_i$ is force on the atom 
at the $i^{th} $ time step.  The force field experience by an atom in the cell will consist of three parts. The deterministic force, 
arising from the nearest neighbor interaction potential, and it plays the role of the external force $F(t)$ acting on the atom. There 
will be energy exchange between the atom and the surrounding atoms in a result of which the atom loses a part of its KE for
 exciting various degrees of freedom of the atoms in the ensemble as well there will be increase in the PE in the cost of KE 
of the atom. This can be described with help of a frictional force $ F_f(t)$ acting on the atom. Besides loosing KE due to frictional effect there is a possibility of gaining KE in the form of random movement of the
 atom due to interactions from all other atoms in the cell and can be modeled by a random force $\Gamma(t) $ acting on the atom. 
$\Gamma(t) $  will have
 the property as  $\langle\Gamma(t)\rangle=0 $   and $\langle\Gamma(t_1)\Gamma(t_2)\rangle=Ag(\Delta t) $  where
 $\Delta t = (t_2-t_1)$, the function $g(t)$ dies down rapidly with $t$ and $A$ is the strength of the random force 
and is a function of temperature T. Thus the dynamics of the atoms will follow Langevin type 
equation \cite{23}. To see the effect of this random force field on the scaling behavior, the ensembles of Si and Ge atoms were exited
 to a higher temperature i.e. at 800 K and allowed to equilibrate at 450 K. The  $\delta $ and $H$ values for the time series of the KE for
 these cases are given in Table~\ref{table2}.  As the systems were allowed to equilibrate of its own without any defined protocol, forces 
acting on the atoms are time dependent through the dynamical process of equilibration. The force field experienced by an atom in the 
\begin{figure}[!ht]
\includegraphics[width= 0.45\textwidth]{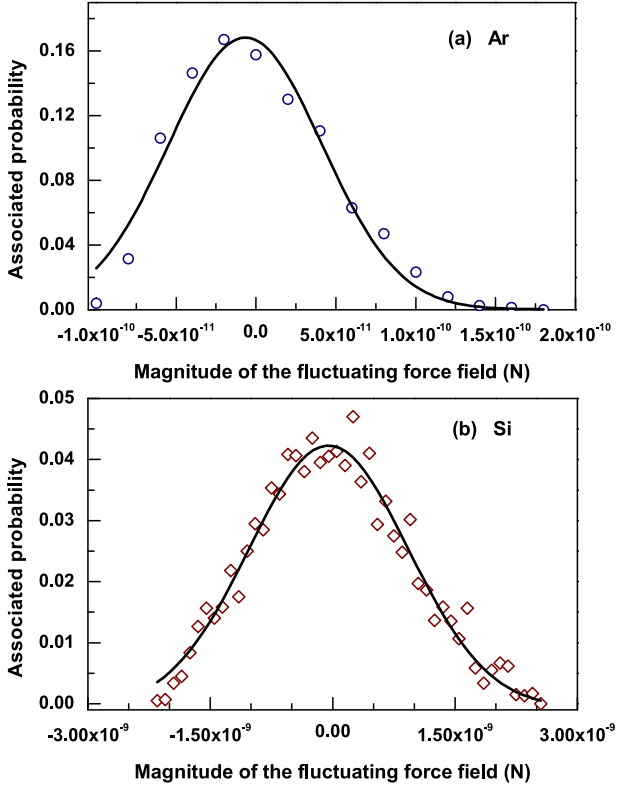}
\caption{Probability distributions of the fluctuating forced fields experienced by (a) an Argon and (b) a Silicon  atom when brought to a 
non-equilibrium state at 500 K from 100 K and allowed to equilibrate at 300 K}
\vspace{-8pt}
\label{figg3}
\end{figure}
simulation cell in the $i^{th}$ time step is calculated from its velocities, at $i-1$, $i$, $i+1^{th}$ time steps. Typical mean values of 
these forces are $1.011654\times10^{-10}$ N and $24.7088\times10^{-10}$ N for solid Argon and Si respectively. The fluctuating
 part of the force field at different time steps was obtained by subtracting the mean values. Typical probability distributions for the
 fluctuating part of the force field in case of solid Ar and Si atoms are shown in Fig.~\ref{figg3}.
 The distributions are Gaussian with
 center $-6.69 \times 10^{-12}$ N and width $9.59\times 10^{-11}$ N for Ar and the corresponding values for Si are
 $-4.98\times10^{-11}$ N and $1.95 \times 10^{-9}$ N respectively. To understand the correlation of the fluctuating force field, 
vectors $ \{\lvert X_i\rangle\}$ are generated from the time series data of the fluctuating force field by taking consecutive n
 time steps for constructing each vector. The eigen value spectrum
 of the covariance matrix $ \Sigma_x $
 of these vectors are calculated to establish the nature of the correlation. Fig.~\ref{figg4} shows typical eigen
 value spectra for Ar and Si cases. The flatness of the spectra confirms that the fluctuating force fields are uncorrelated. 

The motion of the atoms in the simulation cell is thus governed by these forces and dictates the variation of KE of the atoms. 
 The memory effect of the initial velocity will die down with time. In the long time limit the system equilibrates leading to
 equipartition of energy. The characteristic of the variation of KE in terms of time series indicates the evolution of the 
non-equilibrium state. To understand the universality of this evolution for different dynamical processes as generated by MD
 simulations, the Shannon entropy of the diffusion process of KE of an atom in the cell was obtained by calculating the PDF from
 the sub trajectories of this time series. The high values of $\delta$ and $H$ as given in Table~\ref{table2} signifies a strong persistence in the 
fluctuations of the KE of the atoms. The values of $H$ are always larger than that of $\delta$ for all cases studied and are seen to 
fulfill the Levy-walk diffusion relation within the error bar as shown in Table~\ref{table2}.

The exact inherent dynamics of the process of transmitting excess KE of an atom to its surroundings could not be revealed from this
 analysis. However, this analysis reflected the underlying generic features and physical principles that are independent of the
 detailed dynamics or characteristics of particular model. The diffusion of KE of an atom to its surroundings is a continuous 
stationary stochastic process as the probability of the diffusion trajectories follows a scaling relation. As the exponent $H$ is 
greater than 0.5 for all cases, the diffusion is anomalous super diffusion. Any diffusion is a kind of random walk and the Levy-walk
 is a mathematical model to describe anomalous super diffusion where the scaling exponent of variance against time is greater than 
one. Levy-walks have coupled space-time probability distributions and are characterized by a cluster of smaller variations of the
 random variable with a few large variations between them. This pattern repeats for all scales. As the process of equilibration of
 KE of an atom in the ensemble from its non-equilibrium state has to be very rapid, it cannot be Brownian type and it should be 
Levy-walk type as Levy-walk will outperform Brownian walk during the process of equilibration. The Levy-walk type of diffusion 
of KE among the atoms may be due to the following reason. Excess KE of an atom should disburse locally however there is a finite
 probability to transport its excess KE to a distant atom and consequently this atom again redistributes its KE in the same process. 
This process is much faster than normal diffusion when the mean squared value of fluctuation depends on t only. \revision{Mostly three dimensional systems show normal diffusion except in glassy systems \cite{24}. The anomalous or superdiffusion and Levy-walks have been observed in various real-life phenomena like fluid flow in rotating annulus \cite{25}, low dimensional heat transport \cite{26}, light scattering in porous media \cite{27} etc. However our findings show that the diffusion of  KE for a single atom in its non-equilibrium state, when embedded in an environment of atoms that are either in equilibrium or in non-equilibrium state, show superdiffusion and Levy-walk properties throughout the process of equilibration.} In the case when a single atom interacts with the ensemble of atoms that are in equilibrium we can assume 
that the atom is interacting with a thermal bath by absorbing or releasing KE without appreciable change in the bath state.
 In this case also we see the diffusion of KE from the atom is a Levy-walk process. 

Atoms of crystals when brought to a non-equilibrium state, try to redistribute its energy amongst them to converge to an 
equilibrium distribution. This fundamental process is governed by the increase in entropy of the trajectories of the individual 
atom and should be identical in nature for all types of elemental atoms and will be independent of the nature of interactions 
between the atoms and the perturbation which brought them to non-equilibrium state. The calculation of Shannon entropy of the 
diffusion process was made for one atom. However, the concept of entropy in statistical mechanics is for an ensemble. But the
 entropy production for a single trajectory has been addressed in the literature \cite{8} concerning fluctuation theorems.

Atoms in a lattice possessing KE different from its equilibrium values at any instant of time try to transport or accept from 
the nearest neighbors and arrive at a new value. The time series of this KE were translated in to a diffusion process in the
 form of diffusion trajectories and PDF of these trajectories were estimated. The estimated Shannon entropy production of this
 dynamical process with time is found to obey a universal scaling relation. This scaling relation is exact and valid for systems 
no matter how far they are driven out of equilibrium and is independent of the strength of perturbation that brought the system
 out of equilibrium.

\end{document}